\documentclass[letterpaper,11pt]{article}
\usepackage{jheppub}
\usepackage{physics}
\usepackage{relsize}
\usepackage{tikz}
\usepackage{pgfplots}
\usepackage{comment}

\usepackage[utf8]{inputenc}

\title{
Quantum corrections to finite radius holography and holographic entanglement entropy
}

\author[a]{William Donnelly,}
\author[a]{Elise LePage,}
\author[a]{Yan-Yan Li,}
\author[a,b]{Andre Pereira,}
\author[a]{Vasudev Shyam}
\emailAdd{wdonnelly@perimeterinstitute.ca}
\emailAdd{elepage1@perimeterinstitute.ca}
\emailAdd{yli1@perimeterinstitute.ca}
\emailAdd{apereira1@perimeterinstitute.ca}
\emailAdd{vshyam@perimeterinstitute.ca}

\affiliation[a]{Perimeter Institute for Theoretical Physics, \\ 31 Caroline St. N, N2L 2Y5, Waterloo ON, Canada}
\affiliation[b]{Instituto de Física Teórica, UNESP-Universidade Estadual Paulista \\ R. Dr. Bento T. Ferraz 271, Bl. II, São Paulo 01140-070, SP, Brazil}

\abstract{
We calculate quantum corrections to holographic entanglement entropy in the proposed duality between $T\bar{T}$-deformed holographic 2D CFTs and gravity in AdS$_{3}$ with a finite cutoff.
We first establish the dictionary between the two theories by mapping the flow equation of the deformed CFT to the bulk Wheeler-DeWitt equation.
The latter reduces to an ordinary differential equation for the sphere partition function, which we solve to find the entanglement entropy for an entangling surface consisting of two antipodal points on a sphere.
The entanglement entropy in the inverse central charge expansion yields the expectation value of the bulk length operator plus the entropy of length fluctuations, in accordance with the Ryu--Takayanagi formula and its generalization due to Faulkner, Lewkowycz, and Maldacena.
Special attention is paid to the conformal mode problem and its resolution by a choice of contour for the gravitational path integral.
}

\begin{document}

\maketitle

\section{Introduction}
A crucially important question for holography is how to define quantum gravity in regions of finite volume. 
Progress in this issue opens the door to applying holography to questions of quantum cosmology and could provide a more direct route to tackling the paradoxes associated to black hole information loss.
From the quantum field theory perspective, since the cutoff scale of the theory living on the boundary is related to the radius at which the boundary is situated in the bulk, having a dictionary for regions of finite radius could potentially lead to applying holographic methods to strongly coupled quantum field theories possessing some finite scale, i.e. away from criticality.  

In recent years, due to McGough, Mezei and Verlinde \cite{McGough:2016lol}, a concrete proposal has emerged for what theory lives on such a finite cutoff surface in AdS$_{3}$ in the truncation where the effective field theory in the bulk theory is just general relativity. It was also argued that the flow triggered by this deformation on CFTs of large central charge manifests emergent bulk diffeomorphism invariance \cite{Shyam:2017znq}. 
This proposed dual theory is the so-called $T\bar{T}$ deformation of the holographic CFT, introduced in \cite{Smirnov:2016lqw,Cavaglia:2016oda}. 

The $T\bar{T}$ deformation is a solvable irrelevant deformation of two dimensional quantum field theories by an operator quadratic in the stress tensor. 
This deformation has many interesting properties: it preserves integrability when applied to integrable quantum field theories \cite{Smirnov:2016lqw}, and the partition function of such theories obeys a stochastic differential equation and can be thought to arise from turning on Gaussian background metric fluctuations for the undeformed theory \cite{Cardy:2018sdv}. 
A similar proposal was made in  \cite{Dubovsky:2018bmo,Dubovsky:2017cnj}, where the partition function of the deformed theory at any value of the deformation parameter is given by coupling the undeformed theory to Jackiw--Teitelboim gravity. 
This procedure was used to compute the partition function of such theories on a torus.
This torus partition function has also been shown to manifest certain unique modular invariance properties \cite{Datta:2018thy,Aharony:2018bad}.  
The holographic proposal has been subjected to various checks including correlation functions of local operators \cite{Kraus:2018xrn}, propagation speeds and matching of appropriately defined quasilocal energy in the bulk to the ground state of the boundary field theory on a cylinder \cite{McGough:2016lol}. Higher dimensional generalizations of this operator were studied in \cite{Taylor:2018xcy}, \cite{Hartman:2018tkw},\cite{Shyam:2018sro} and in particular, the sphere partition function at large $c$ in dimensions up to 6 was computed in \cite{Caputa:2019pam} and the entanglement entropy of the Hartle Hawking state were extracted from these results in \cite{Banerjee:2019ewu}, \cite{Murdia:2019fax}.

Here we will be interested in the entanglement entropy of the $T\overline{T}$-deformed CFT and its relation to the holographic entanglement entropy.
The entanglement entropy for an entangling surface consisting of two antipodal points on a sphere was calculated in the large $c$ limit in Ref.~\cite{Donnelly:2018bef}, and it was shown to satisfy a finite-cutoff version of the Ryu--Takayanagi formula \cite{Ryu:2006bv}.\footnote{The conical entropy, a close relative of the R\'enyi entropy, was also calculated and found to give a finite-cutoff version of a formula due to Dong \cite{Dong:2016fnf}.} 
Our goal in this work is to study corrections to these large $c$ results, which are expected to capture quantum corrections to the bulk gravitational theory.

We begin in section \ref{section:ttbar} with the definition of the $T\bar{T}$ flow equation on $S^{2}$ at finite central charge.
We find that it gives the Wheeler-DeWitt equation for the wavefunctional of the metric; this fixes the relation between the boundary constants (central charge $c$ and $T \overline T$ deformation parameter $\mu$) and bulk constants (Newton's constant $G$ and AdS radius $\ell$) as well as the relation between the boundary partition function and the bulk wavefunction.
This generalizes the argument of Ref.~\cite{McGough:2016lol} to finite $c$, and does not rely on the assumption of factorization.
Using methods of canonical quantum gravity we reduce the phase space to the spherically symmetric sector, where the Wheeler-DeWitt equation becomes an ordinary differential equation.

In section \ref{section:Z} we consider the solution of the Wheeler-DeWitt equation.
While the equation admits an exact solution in terms of confluent hypergeometric functions, we first find the partition function using the WKB method which allows us to express the result as an expansion around the classical saddle point, which matches the classical action of AdS${}_3$ with a radial cutoff.
To find the exact solution we have to specify boundary conditions in the small $r$ regime, which is outside the regime of validity of the WKB approximation.
The resulting ambiguity corresponds to an ambiguity of choice of contour in the path integral.
While there is no universally agreed upon choice for the contour for a gravitational path integral, by a combination of a choice of contour and a boundary condition we are able to fix the partition function in a way that picks out the correct classical saddle point in the classical regime while having sensible behaviour in the deep ultraviolet.

In section \ref{section:entanglement} we apply these considerations to entanglement entropy, following the setup of \cite{Donnelly:2018bef}.
In the semiclassical expansion, we obtain the entanglement entropy as the length of a bulk geodesic plus quantum corrections.
Using the path integral expression for the partition function, we can express this result nonperturbatively as the expectation value of the length operator, plus a term which captures the entropy of fluctuations of the length.

\section{$T\bar{T}$ deformation and finite radius holography}

\label{section:ttbar}

The operator $T\bar{T}$ is defined as the coincidence limit of the bilocal operator 
\begin{equation}
    T\bar{T}(x)=
8 \, G_{ijkl}(x) \, \lim_{y\rightarrow x}  \left( T^{ij}(x)\,T^{kl}(y)\right),
\end{equation}
where we have introduced the \emph{DeWitt supermetric} $G_{ijkl}(x)$ constructed from the two-dimensional metric $\gamma_{ij}(x)$:
\begin{equation}
    G_{ijkl}(x) \equiv \gamma_{i(k}(x)\gamma_{l)j}(x)-\gamma_{ij}(x)\gamma_{kl}(x).
\end{equation}
The DeWitt supermetric appears naturally in the Hamiltonian formulation of gravity as the coefficient of the kinetic term for the metric. 
Writing the $T \bar T$ operator in terms of $G_{ijkl}$ makes the connection to gravity manifest, and also leads to a natural generalization to higher dimensions.

The starting point of our analysis will be the flow equation defining the $T\bar{T}$ deformation of a two-dimensional conformal field theory with central charge $c$. 

We will assume that the only deforming operator is $\mathcal{O} = T\bar{T}$, with coupling constant $\mu$.
There is only one dimensionful scale in the theory: it is defined by $\mu$, which has dimensions of length squared. This means that the expectation value of this operator can be obtained by taking a derivative with respect to $\mu$:
\begin{equation}
\partial_{\mu}Z=\frac{1}{4}\int \langle T\bar{T}(x)\rangle Z. \label{cflowe}
\end{equation}
We will promote the deformation parameter $\mu$ to a function $\mu \lambda(x)$ and ask what happens when we compute the trace of the stress tensor. In other words, that the equation \eqref{cflowe} can be upgraded to 
\begin{equation}
\frac{\delta Z}{\delta \lambda(x)}=\frac{\mu}{4} \langle T\bar{T}(x)\rangle Z.
\end{equation}
Recalling that the trace of the stress tensor encodes the response of the partition function under the change of every length scale present in the theory:
\begin{equation}
\gamma_{ij}\frac{\delta Z}{\delta \gamma_{ij}}=\langle T^{i}_{i}(x) \rangle=-\mu\frac{\delta Z}{\delta \lambda(x)}-\frac{c}{24\pi}R(x),
\end{equation}
we obtain the flow equation:
\begin{equation}
 \ev{ T^{i}_{~i}(x)}=-\frac{\mu}{4}  \ev{ T\bar T(x) } -\frac{c}{24\pi}R(x).\label{fe}
\end{equation}
We assume that no other scale is generated, and hence that we can extrapolate \eqref{fe} to finite $\mu$.

Recall that the partition function, viewed as a functional of the metric, is a generating functional for stress-tensor correlation functions:
\begin{equation} \label{Tcorrelator}
  \ev{ T^{ij}(x_1) \cdots T^{kl}(x_n) } = \frac{1}{Z[\gamma]}
  \left( \frac{-2}{\sqrt{\gamma(x_1)}} \frac{ \delta}{\delta \gamma_{ij}(x_1)} \right) \cdots \left( \frac{-2}{\sqrt{\gamma(x_n)}} \frac{ \delta}{\delta \gamma_{kl}(x_n)} \right)
  Z[\gamma].
\end{equation}
Substituting \eqref{Tcorrelator} into the flow equation \eqref{fe} yields a functional differential equation for the partition function $Z[\gamma]$:
\begin{equation} \label{fez}
 -\frac{2}{\sqrt{\gamma(x)}}\gamma_{ij}(x) \frac{\delta Z[\gamma]}{\delta \gamma_{ij}(x)}=-\lim_{y\rightarrow x}\frac{\mu \, G_{ijkl}(x)}{\sqrt{\gamma(x)}\sqrt{\gamma(y)}} \, \frac{\delta^{2}Z[\gamma]}{\delta \gamma_{ij}(x)\delta\gamma_{kl}(y)}-\frac{c}{24\pi}R(x) \, Z[\gamma].
\end{equation}
This is a linear second-order functional differential equation for $Z[\gamma]$, a functional of the two-dimensional metric $\gamma$.

\subsection{Flow equation and Wheeler-DeWitt equation
}

In Ref.~\cite{McGough:2016lol}, the $T \bar T$-deformation of conformal field theory at large central charge $c$ was argued to be dual to AdS${}_3$ gravity with a finite-sized boundary.
Here we review this connection with an eye toward extending it beyond the classical limit $c \to \infty$. 
We will first describe how the flow equation at large and finite $c$ map to the Hamiltonian constraint and Wheeler-DeWitt equation in AdS$_{3}$ respectively on general backgrounds. Then, we will specialize to the case where the $T\bar{T}$ deformed theory lives on a spherical background. 

\subsubsection{Large $c$: Hamiltonian constraint}

We first briefly review the derivation, following Ref.~\cite{McGough:2016lol}, of the classical Hamiltonian constraint from the $T \overline{T}$ flow equation under certain assumptions about the factorization properties of the $T \overline T$ operator.

The expectation value of $T\bar{T}(x)$ factorizes as follows in both translation invariant states \cite{Zamolodchikov:2004ce}, and in the limit of the large central charge $c$ on general curved backgrounds \cite{Jiang:2019tcq}:
\begin{equation}
    \ev{T \bar T(x)}  =\ev{T^{ij}(x)}\ev{  T_{ij}(x)}-\ev{ T^{i}_{~i}(x)}^{2}. \label{fac}
\end{equation}
Now, consider the flow equation defining the $T\bar{T}$ deformed CFT at large $c$, where \eqref{fac} holds. 
We then identify the conjugate momentum to the metric
\begin{equation}
    \pi^{ij}=\sqrt{\gamma}\left(\ev{T^{ij}}+\frac{2}{\mu}\gamma^{ij}\right),
\end{equation}
and make the identifications of constants 
\begin{equation}  \label{constants}
\mu=16\pi G \ell, \qquad c=\frac{3\ell}{2G},
\end{equation}
where $G$ is Newton's constant and $\ell$ the AdS radius in the bulk.
Then \eqref{fe} with this factorization becomes the bulk Hamiltonian constraint for gravity with a negative cosmological constant $\Lambda = -\frac{1}{\ell^2}$:
\begin{equation}
\frac{16\pi G}{\sqrt{\gamma}}\left(\pi^{ij}\pi_{ij}-(\pi^{i}_{~i})^{2}\right)+\frac{\sqrt{\gamma}}{16\pi G}\left(R+\frac{2}{\ell^{2}}\right)=0.\label{hc}
\end{equation}
The stress tensor conservation $\nabla_i \ev{T^{ij}} = 0$ becomes the momentum constraint $\nabla_{i}\pi^{ij}=0$.

\subsubsection{Finite $c$: Wheeler-DeWitt equation}

In fact the preceding argument generalizes naturally to finite $c$, without the assumption of factorization \eqref{fac}.

We first introduce a change of variables to eliminate the linear term in \eqref{fez}.
This is given by
\begin{equation}\label{zpsi}
 \Psi[\gamma]=e^{\frac{2}{\mu}\int \textrm{d}^{2}x \sqrt{\gamma}}Z[\gamma].
\end{equation}
\eqref{fez} then becomes the following equation for $\Psi[\gamma]$: 
\begin{equation}
\frac{16\pi G}{\sqrt{\gamma}(x) \sqrt{\gamma}(y)}G_{ijkl}(x)\lim_{x\rightarrow y}:\frac{\delta^{2}\Psi[\gamma]}{\delta \gamma_{ij}(x)\delta \gamma_{kl}(y)}:+\frac{1}{16\pi G}\left(R+\frac{2}{\ell^{2}}\right)\Psi[\gamma]=0.\label{wdw}
\end{equation}
This equation is the radial Wheeler-DeWitt equation in a space with negative cosmological constant provided we identify bulk constants as in \eqref{constants}.
Eq.~\eqref{wdw} can be obtained from the classical constraint equation \eqref{hc} by replacing the momenta with functional derivatives with respect to $\gamma_{ij}(x)$. 
Conversely, the classical constraint equation \eqref{hc} can be obtained from the Wheeler-DeWitt equation \eqref{wdw} in the leading-order WKB approximation, as we will see in detail in the following section. 

The symbol $:(\cdots):$ denotes a procedure for regularizing the functional derivative at coincident points, which we will leave unspecified as it is unnecessary for the purposes of this work.
This is because we will not attempt to solve the above second order functional differential equation.
Instead we will specialize to the case where the constant radius slices are spheres and consider the quantization of the Hamiltonian constraint \eqref{hc} after phase space reduction. 

\subsection{Phase space reduction for $S^{2}$ radial slices}\label{section:s2}

From the point of view of quantizing the bulk theory, it is natural to introduce a gauge fixing for the Hamiltonian constraint.
We choose the constant mean curvature gauge 
\begin{equation}
\frac{\pi^{i}_{i}}{\sqrt{\gamma}}=\tau,
\end{equation}
where $\tau$ is constant. 
We then decompose the momentum conjugate to the metric as follows:
\begin{equation}
\pi^{ij}=\left(\tfrac12 \tau \sqrt{\gamma} \gamma^{ij}+\sigma^{ij}\right),
\end{equation}
where $\sigma^{ij}$ is traceless. 
Imposing the constancy of $\tau$ implies additionally that 
\begin{equation}
\nabla_{i}\sigma^{ij}=0,
\end{equation}
and therefore $\sigma^{ij}$ is transverse and tracefree. 
On the sphere, such tensors must vanish identically, so $\sigma^{ij} = 0$. 
We can then do a conformal decomposition of the metric
\begin{equation} \label{gamma}
\gamma_{ij}=e^{2\lambda(x)}h_{ij},
\end{equation}
where $h_{ij}$ is the standard round metric on the unit 2-sphere, with $R[h] = 2$. 
The Ricci scalar of $\gamma$ is $R[\gamma] = e^{-2 \lambda} (R[h] - 2 \nabla^2 \lambda)$.
Having fixed the gauge as in \eqref{gamma} the Hamiltonian constraint for sphere radial slices becomes an equation for the conformal factor $\lambda$:
\begin{equation} \label{york}
 \Delta \lambda=\frac{R[h]}{2}-e^{2\lambda}\left(\frac{(16\pi G)^{2}\tau^{2}}{4} - \frac{1}{\ell^2}\right).
\end{equation}
This equation has a unique solution up to a zero mode, which was shown in \cite{doi:10.1063/1.528475,10.1143/PTP.83.733} (although for a different combination of signs).
All two dimensional spherical geometries are conformal to the round sphere and solutions to \eqref{york} give us the factor with which to perform the Weyl transformation between the given metric and the round two sphere metric. 
Only the global part of the Hamiltonian constraint remains unfixed. 
This is obtained by integrating the above equation over the sphere  
\begin{equation}
V\left((16\pi G)^{2}\tau^{2}-\frac{2}{\ell^{2}}\right)-4\pi=0,
\end{equation}
where $V = \int d^2x \, \sqrt{h} \, e^{\lambda(x)}$ is the volume. 

York's method involves taking the mean curvature $\tau$ to be `time' and treating the volume as a true Hamiltonian 
\begin{equation}
V(\tau)=\frac{4\pi}{(16\pi G)^{2}\tau^{2}-\frac{2}{\ell^{2}}}.\label{glob}
\end{equation}
This is the well-known deparameterization of the Hamiltonian constraint in terms of York time. 
However it is not convenient for our purposes, since we are interested in the wavefunction of a spherical geometry as a function of radius.
In this representation, the volume is a function of the configuration variable $r$, $V=4\pi r^{2}$.
The mean curvature $\tau$ is the conjugate momentum to the volume, which is related to the momentum conjugate to $r$ as
\begin{equation}
p_{V}=\frac{p_{r}}{8\pi r}.
\end{equation}
With these variables, the classical constraint equation \eqref{glob} becomes
\begin{equation} \label{Hconstraint}
 G^{2}p^{2}_{r}-\left(1+\frac{ r^{2}}{\ell^{2}}\right)=0. 
\end{equation}
The quantization of the above constraint equation is therefore a time-independent Schr\"odinger equation, and the phase space reduction necessitates limiting our attention to the global modes of the geometry. 
We choose to parameterize these modes through the radius $r$ and its conjugate momentum $p_r$. 

\subsection{Symmetry reduced action and Wheeler-DeWitt equation}

In this section we give an alternative derivation of the Wheeler-DeWitt equation \eqref{WDW} by symmetry reduction of the action for AdS${}_3$ gravity, c.f. \cite{Caputa:2018asc}.
This will be useful in order to make contact with the Euclidean path integral formalism in section \ref{section:Z}.

The action for general relativity with a negative cosmological constant including the counterterm \cite{Balasubramanian:1999re} takes the form $S_{GR}=S_{\text{EH}}+ S_{\text{GHY}}+S_{\text{CT}}$ where the various terms are given by:
\begin{align}
 S_{\text{EH}} &= -\frac{1}{16 \pi G} \int \dd^3 x \sqrt{g} \left( R + \frac{2}{\ell^2} \right), \\
 S_{\text{GHY}} &= \frac{1}{8 \pi G} \oint \dd^2 x \sqrt{\gamma} \, K \\
 S_{\text{CT}} &= \frac{1}{8 \pi G \ell} \oint \dd^2 x \sqrt{\gamma}.
\end{align}

Note that the counterterm is precisely the factor appearing in \eqref{zpsi} that relates the deformed CFT partition function and the bulk gravity wavefunction:
\begin{equation}
\Psi[\gamma] = e^{S_\text{CT}} Z[\gamma].
\end{equation}
The context, however, is slightly different from  Ref.~\cite{Balasubramanian:1999re}.
There, the counterterm was required to obtain a finite partition function in the limit where $\gamma$ is large.
Here it appears as a generating function in a canonical transformation that eliminates the first derivative term from the flow equation \eqref{fez}.

We wish then to pass to the Hamiltonian formalism, which first requires foliating the spacetime by hypersurfaces.
We will assume spherical symmetry, under which a general metric takes the form
\begin{equation} \label{frw}
 \dd s^{2} = N^{2}(\rho) \, \dd \rho^{2} + r^{2}(\rho) \, \dd \Omega_{2}.
\end{equation}
This form is quite familiar from studies of homogeneous and isotropic cosmology: the function $r(\rho)$ controls the size of the sphere of fixed $\rho$ and is analogous to the scale factor in the Friedman-Robertson-Walker metric.
The function $N(\rho)$ is analogous to the lapse, the difference being that in the metric \eqref{frw} the normal to the surfaces of constant $\rho$ is spacelike.

The action, not including the counterterm then becomes:
\begin{equation}
 S_\text{EH} + S_\text{GHY} = -\frac{1}{2 G} \, \int \dd \rho ~ N(\rho) \, \left( 1 + \left(\frac{r^{\prime}(\rho)}{N(\rho)}\right)^{2} + \frac{r(\rho)^{2}}{\ell^{2}} \right).
\end{equation}
We note that the Euclidean action is both negative and unbounded below.
The fact that the action is negative is an important feature --- we will see that it is precisely this feature that leads to the positivity of the entropy, and consistency with the Ryu-Takayanagi formula when evaluated on the classical solution.
The fact that the fluctuations around the classical solution also have negative Euclidean action is a serious problem, since the integral of $e^{-S}$ will diverge.
This is the famous conformal mode problem, which we will be forced to revisit in section \ref{section:Z}.

The derivation of the Wheeler-DeWitt equation from the symmetry reduced action is standard.
We first identify the conjugate momenta to $r$ and $N$:
\begin{equation}
    p_r = - \frac{r'}{GN}, \quad p_N = 0.
\end{equation}
The latter is a constraint, and its preservation leads directly to the Hamiltonian constraint
\begin{equation}
    G^2 p_r^2 - \left( 1 + \frac{r^2}{\ell^2} \right) =0.
\end{equation}
This agrees with the result \eqref{Hconstraint} derived from gauge-fixing in the hamiltonian formalism.

The classical limit in the bulk theory is one where $G \ll \ell$, which in terms of the field theory implies that $c\gg1$. 
In that limit the $T\overline{T}$ flow equation becomes the radial Hamiltonian constraint in the bulk. 
In the minisuperspace approximation, i.e. when we truncate to the symmetry reduced sector, this constraint reduces to \eqref{Hconstraint}. 
When we quantize this theory, we obtain an equation valid when $G$ and $\ell$ are comparable, which translates into $c$ remaining $\mathcal{O}(1)$. 
This is why the arbitrary $c$ flow equation in the symmetry reduced $T\bar{T}$ theory should be identified with the minisuperspace Hamiltonian constraint.

\subsection{Emergent diffeomorphism invariance}
The phase space reduction presented in \ref{section:s2} shows us how the large $c$ flow equation is written purely in terms of variations with respect to the global geometrical modes. From the intrinsically two dimensional point of view, this is to be expected given that the flow equation is written in terms of only one point functions of the stress tensor which are themselves given purely in terms of the derivatives of the partition function with respect to the radius. 

However, the fact that the RG flow equation even at finite $c$ involves only derivatives with respect to the global modes of the metric is nontrivial. This happens for the $T\bar{T}$ flow equation (in \cite{Dubovsky:2018bmo}, \cite{Cardy:2018sdv}) on $\mathbb{T}^{2}$ as well, except for very different reasons \footnote{The $T\bar{T}$ flow equation is the one that reads $\partial_{\mu}Z=\langle T\bar{T}(x)\rangle Z $, and it tells us how the quantum field theory responds under the change of one scale in the problem, i.e. the one associated to the $T\bar{T}$ operator. The RG flow equation or Callan--Symanzik equation on the other hand tells us how the theory responds to a local change of scale}. There, the localisation of the flow equation on to the zero mode sector is due to the separation independence of the contracted two point function of stress tensors whose coincidence limit defines the $T\bar{T}$ operator \cite{ZoharBootstrap}. 
 
In our case however, it is crucially important that the flow equation that arises from deforming a conformal field theory with $T\bar{T}$ can be rewritten as the Wheeler-DeWitt equation. 
The Wheeler-DeWitt equation and the Ward identity $\nabla_{i}\langle T^{ij}\rangle=0$ encode the invariance of the wave function $\Psi[\gamma]$ under normal and tangential deformations of the hyper surface on which it is evaluated. 
In quantum gravity, these deformations describe the action of diffeomorphisms of the bulk space time into which the hypersurface is embedded.

One could ask how important it was to deform a holographic CFT in order to exploit bulk diffeomorphism invariance. We argue that in fact it isn't important, since all we have done is to make a change of variables and identified constants in a certain manner. When we are considering pure gravity in the bulk, this agrees with a finite cutoff generalization of the conventional AdS/CFT dictionary. However, when other matter fields are involved in the bulk, in order to maintain the identifications as dictated by the AdS/CFT dictionary, this mapping is inadequate \cite{Kraus:2018xrn} and other double trace deformations involving operators other than the stress tensor must be included in the boundary theory \cite{Hartman:2018tkw}.

Given the expectation that correlation functions of local operators are expected to be smeared or delocalized by the $T\bar{T}$ deformation \cite{Cardy:2019qao}, even if the standard dictionary is maintained, unless other double-trace deformations are included, the bulk theory likely also involves nonlocal matter fields.

However, one can also consider the stress tensor sector of a general CFT at finite $c$, which isn't expected to possess a classical bulk dual. 
In such a theory, we can apply the $T\bar{T}$ deformation and find that the local RG flow equation will take the form \eqref{fez}.
After making some simple identification of the constants and by redefining the partition function in terms of $\Psi[\gamma]$, this flow equation takes the form of the Wheeler-DeWitt equation \eqref{wdw}, irrespective of whether or not it has a semiclassical bulk dual. \footnote{We acknowledge that Aitor Lewkowycz independently realised this perspective on $T\bar{T}$ deformed theories}

In this article, we are interested in computing the sphere partition function in a $T\bar{T}$ deformed CFT with some arbitrary, finite central charge. Although the method involves exploiting the emergent bulk diffeomorphism invariance in order to turn our problem into an effectively quantum mechanical one, we do not require the theory we are deforming to possess a holographically dual description in terms of string theory on AdS$_{3}$.

\section{Sphere partition function}
\label{section:Z}

The quantization of the reduced phase space Hamiltonian constraint leads to the following spherically symmetric Wheeler-DeWitt equation:
\begin{equation} \label{WDW}
G^2 \left( \dv[2]{r} +\frac{2b-1}{r}\frac{\textrm{d}}{\textrm{d}r}\right)\psi(r) 
 = \left( 1 + \frac{r^{2}}{\ell^{2}} \right) \, \psi(r),
\end{equation}
where $\psi(r)=\Psi[\gamma_{S^{2}}]$ is the wavefunctional evaluated on a sphere of radius $r$. 
The constant $b$ appears due to the ordering ambiguity for the kinetic term. 
Then, if we recall the relationship between the partition function and the solution to the Wheeler-DeWitt equation \eqref{zpsi}, the sphere partition function of the $T\bar{T}$ deformed CFT is given by 
\begin{equation}
Z(r) = e^{-\frac{r^2}{2 G \ell}}
\psi(r).
\end{equation}
where we have used the identifications \eqref{constants}.

It will be convenient to work in terms of gravitational units and set $\ell = 1$.
In these units $G$ becomes a dimensionless parameter, the ratio of the Planck length to the AdS radius, which is small in the classical limit.
Although the equation admits an exact solution in terms of special functions, it will be instructive to first study the solution semiclassically.

\subsection{WKB approximation}

When $G$ is small, the equation \eqref{WDW} can be treated by the WKB approximation.
It will be convenient to define $\psi = e^W$, where $W = \log Z + S_\text{CT}$ is the effective action, up to the counterterm.
We then expand in powers of $G$,
\begin{equation}
W = \frac{1}{G} W_0 + W_1 + G W_2 + \ldots.
\end{equation}
In terms of the effective action, equation \eqref{WDW} becomes
\begin{equation}
G^2 \left( W'' + (W')^2 + \frac{2b-1}{r} W' \right) = (1 + r^2).
\end{equation}
where $'$ denotes $\dv{r}$.

For a second order equation, there are two classical solutions $W_0^\pm$.
We can then consider the expansion around each of these solutions, and a general solution is given by:
\begin{equation}
\psi(r) = \alpha_+ e^{\frac{1}{G} W_0^+ + W_1^+ + \ldots} + \alpha_- e^{\frac{1}{G}W_0^- + W_1^- + \ldots}
\end{equation}
The negative solution $W_0^-$ is suppressed relative to the positive solution by the exponential of the classical action. 
This is nonperturbatively small when $r$ is larger than the Planck scale. 
For now we focus on corrections around the dominant saddle point.

\paragraph{Classical solution}
At leading order in the $G$ expansion we have two solutions for $W_0$,
\begin{equation}
W'_0(r) = \pm \sqrt{r^2 + 1}.
\end{equation}
The positive solution is given by
\begin{equation} \label{W0}
    W^+_0(r) = \frac{1}{2} (\sinh^{-1}(r) + r \sqrt{r^2 + 1}).
\end{equation}
This solution corresponds to the Euclidean action (without the holographic counterterm) evaluated on the classical saddle point, which in this case is a region of Euclidean AdS space bounded by a sphere of radius $r$.
Restoring $\ell$ and the counterterm, this corresponds to a classical solution
\begin{equation}
    Z(r) \sim \exp\left( \frac{\ell}{2 G} \left(\sinh^{-1} \left(\frac{r}{\ell} \right) + \frac{r}{\ell} \sqrt{\frac{r^2}{\ell^2}+1} - \frac{r^2}{\ell^2}
    \right)\right)
\end{equation}
This agrees with the evaluation of the classical action for a region of Euclidean AdS${}_3$ bounded by a sphere of radius $r$ \cite{Donnelly:2018bef}.

The other classical solution $W_0^-(r)$ corresponds to the opposite sign of \eqref{W0}.
This yields a saddle point whose contribution to the partition function is exponentially suppressed when $r \gg G$.

\paragraph{One-loop correction}

The WKB expansion also allows us to find subleading corrections in the loop expansion.
Substituting the leading-order WKB solution into the first order equation yields
\begin{equation} \label{WKB1}
W_1'(r) = -\frac{1}{2} \left( \frac{2b-1}{r} + \frac{W_0''(r)}{W_0'(r)} \right).
\end{equation}
The one-loop correction is given by
\begin{equation}
W_1(r) = -\tfrac12 (2b-1) \log(r) - \tfrac14 \log (1+r^2).
\end{equation}
Note that the one-loop correction is the same around both classical solutions $W_0^\pm$, since \eqref{WKB1} is unaffected by flipping the sign of $W_0$.

Note also that the corrections become large as $r \to 0$, while the leading term $W_0(r)$ vanishes in that limit.
This indicates that the WKB approximation breaks down at distances approaching the Planck scale.

From this result, we can infer the leading order quantum corrected partition function in the bulk:
\begin{equation}
    Z(r) \sim \exp\left( \frac{\ell}{2 G} \left(\sinh^{-1} \left(\frac{r}{\ell} \right) + \frac{r}{\ell} \sqrt{\frac{r^2}{\ell^2}+1} - \frac{r^2}{\ell^2}
    \right) + \frac{1}{4} \log \left( \frac{r^2}{r^2 + \ell^2} \right)
     - b \log \left( \frac{r}{\ell}\right)
     \right).
\end{equation}
There are two important caveats with this solution: we have neglected the contribution of the subdominant saddle point, and we have not fixed the constant in front of the solution.
To appropriately resolve this issue we require boundary conditions as $r \to 0$.
Since the point $r = 0$ is outside the regime of validity of the WKB approximation, we have to use other methods to determine the solution in that regime.

\subsection{Exact solution}
\label{subsection:exact}

Equation \eqref{WDW} admits an exact solution.
Changing to the independent variable to $z = r^2/G$ and rescaling $\psi$ to $g$ as
\begin{equation}
g(z) = e^{z/2} \psi(\sqrt{G z}),
\end{equation}
\eqref{WDW} becomes Kummer's equation (\cite{NIST:DLMF}, \S13)
\begin{equation} \label{kummer}
z g''(z) + (b - z) g'(z) - a g(z) = 0,
\end{equation}
where $a = \frac{1}{4G} + \frac{b}{2}$.
When $b \notin \mathbb{Z}$, the general solution of \eqref{kummer} is given by a linear combination of the confluent hypergeometric functions $U(a,b,z)$ and $M(a,b,z)$.\footnote{$M(a,b,z)$ is sometimes denoted ${}_1 F_1(a;b;z)$.}
For $b \in \mathbb{Z}$ the expansion for $z \to 0$ is more complicated and so we will assume the generic case $b \notin \mathbb{Z}$ from here on.

In the limit $z \to 0$, the solution is parameterized by
\begin{equation}
    g(z) = c_1 M(a,b,z) + c_2 z^{1-b} M(a-b+1,2-b,z).
\end{equation}
As $z \to 0$,
\begin{equation}
g(z) = c_1 (1 + O(z)) + c_2 z^{1-b}( 1 + O(z)).
\end{equation}
The constants $c_1,c_2$ are determined from the boundary conditions as $z \to 0$.
In the classical solution \eqref{W0}, we only need a single boundary condition since the leading order WKB equation is first-order.
However, the Wheeler-DeWitt equation is second-order and so we require a second boundary condition to fully specify the solution.
This additional boundary condition determines the contribution from the subdominant saddle point.

The boundary condition chosen in Ref.\cite{Donnelly:2018bef} was to take $Z = 1 + O(r^2)$ as $r \to 0$.
The simplest choice which achieves this is $c_1 = 1$, $c_2 = 0$.
This leads to a partition function that coincides with that of a trivial theory in the ultraviolet, and we will see in section \ref{section:entanglement} that this makes the entanglement entropy vanish as $r \to 0$ as might be expected if the $T \overline{T}$ deformation acts as an effective ultraviolet cutoff.\footnote{An alternative prescription \cite{Gorbenko:2018oov} is to start from the conformal field theory in the limit $r \to \infty$.
In this limit the partition function has an undetermined constant coming from the cutoff scale.}

\subsection{Path integral representation}

The preceding results can also be obtained from a Euclidean path integral.
This will be useful in the following section in comparing the entanglement entropy with the bulk length.

We first deparametrize the system by introducing a parameter $L$ be the diameter of the bulk, i.e. twice the proper distance from the center to a sphere of fixed radius $r$.
The wavefunction $\psi(r)$ is replaced with a wavefunction $\psi(r,L)$ in which $L$ plays the r\^ole of a Euclidean time coordinate.
The Wheeler-DeWitt equation is then replaced with the Euclidean Schr\"odinger equation:
\begin{equation} \label{schrodinger}
    \left(- 4 G \pdv{L} - G^2 \left( \pdv[2]{r} + \frac{2b-1}{r} \pdv{r} \right) + r^2 + 1 \right) \psi(r,L) = 0.
\end{equation}
The solution of the Wheeler-DeWitt equation will be obtained by integration,
\begin{equation} \label{pr}
\psi(r) = \int \mu_L dL \; \psi(r,L).
\end{equation}
The measure factor $\mu_L$ is required by dimensional analysis and has units of inverse length.
We also leave the contour of integration unspecified; provided $\psi(r,L)$ solves \eqref{schrodinger}, then the integral \eqref{pr} will solve the Wheeler-DeWitt equation if the contour is chosen such that the contribution from the endpoints vanishes.

We can solve this with an ansatz $\psi(r,L) = e^{\alpha(L) + \beta(L) r^2}$, which leads to a pair of equations
\begin{align}
    4 G \dv{L}\beta + 4 G^2 \beta^2 - 1  &= 0 \\
    4 G \dv{L}\alpha + 4 b  G^2 \beta - 1 &= 0.
\end{align}
These equations are easily integrated, but two constants of integration must be specified.
One constant can be absorbed into a redefinition of $L$ (and hence in a shift of the contour used in integrating $L$).
The second constant can be absorbed into the measure factor $\mu_L$.
Having made these choices, the solution is given by:
\begin{align}
    \beta &= \frac{1}{2 G } \coth \left( \frac{L}{2} \right) \\
    \alpha &= \frac{L}{4 G} - b \log \sinh \left( \frac{L}{2} \right).
\end{align}
Which yields
\begin{equation} \label{prL}
    \psi(r,L) =  \sinh \left( \frac{L}{2} \right)^{-b} \exp \left[ \frac{L}{4G} + \frac{r^2}{2 G} \coth \left( \frac{L}{2} \right)\right].
\end{equation}

Before carrying out the path integral, we first look at the classical solutions.
The exponential has two real saddle points $\pm L_0$, where
\begin{equation}
 \sinh(L_0/2) = r.
\end{equation}
The positive saddle point at $L = L_0$ corresponds to Euclidean AdS3 with line element
\begin{equation}
    \dd s^2 = \dd \rho^2 + \sinh(\rho)^2 \dd \Omega_2
\end{equation}
which has $r = \sinh(\rho)$ and $L = 2 \rho$.
Evaluating $\psi(r,L)$ at this point yields 
\begin{equation}
\psi(r,L_0) = r^{-b} \exp \left[ \frac{1}{2G} \left( \sinh^{-1}(r) + r \sqrt{r^2 + 1}\right) \right]
\end{equation}
The quantity in the exponential is precisely the classical solution $W_0^+$ given by the WKB method \eqref{W0}.
The saddle point at $L = -L_0$ gives the exponentially suppressed WKB solution $W_0^-$.
\footnote{There are an infinite number of complex saddle points at $L = \pm L_0 + 2 \pi i n$ for $n \in \mathbb{Z}$.
Shifting the imaginary part of $L \to L + 2 \pi i n$ shifts the integrand as $\psi(r,L + 2 \pi i n) = \psi(r,L) e^{ \frac{\pi i n}{2 G}}$, so the complex contours and saddle points differ from their real counterparts by a phase.
}

We now turn to carrying out the Euclidean path integral, and the choice of contour for $L$.
Since $L$ is a Euclidean length, it would seem natural to integrate over positive real $L$.
However, the result \eqref{prL} diverges as $e^{L}$ for large $L$, and as $e^{1/L}$ for small $L$.
This is a manifestation of the conformal mode problem; the Euclidean action we started with is not bounded below.
We note that there is not a single agreed-upon prescription for carrying out gravitational path integrals of this type; but a list of desired criteria were outlined in Ref.~\cite{Halliwell:1989dy}. 

There are other remedies for this issue in the literature in the context of performing the full Euclidean path integral (i.e. beyond symmetry reduction in $d>3$).
One such prescription is to perform a Wick rotation in a `proper time gauge' of the metric fluctuations, as discussed in \cite{Dasgupta:2001ue}. 
Another way to circumvent the issue involves a nonlocal field redefinition \cite{Mazur:1989by}.
In both of these contexts, a Jacobian arising from the path integral measure cancels the divergence of the Euclidean action. 

We know that on general grounds, whatever contour we choose to integrate \eqref{pr}, it should be deformable to a combination of steepest descent contours which will pass through some set of saddle points.
In order to match with the expected classical behavior, this set of saddle points must include the positive saddle $L_0$.
We will also demand that the solution $\psi(r)$ is real:
the original Euclidean integral we want to deform, though divergent, is formally real, and we will see in section \ref{section:entanglement} that complex solutions for $\psi(r)$ lead to complex entropy.

\begin{figure}
    \centering
    \includegraphics{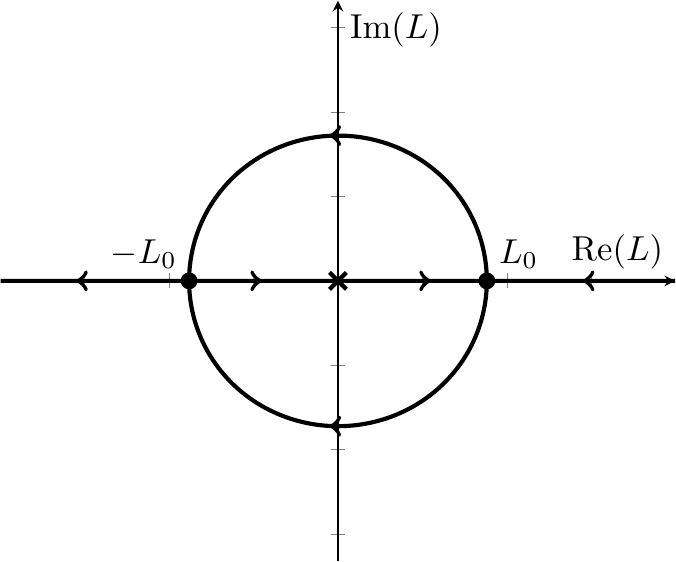}
\caption{Steepest descent contours for the evaluation of the integral of $\psi(r,L)$.
The thick lines denote values for which $\psi(r,L)$ is real; here we have set $r = 1$, but the qualitative behaviour is independent of $r$.
Direction of the arrows denotes the direction in which the integrand is decreasing.
The cross at the origin denotes an essential singularity of the integrand.
}
    \label{fig:contours}
\end{figure}

The steepest descent curves passing through the real saddle points have stationary phase, which means the quantity in the exponential of \eqref{prL} is real. 
Letting $L = x + i y$, these curves are solutions of
\begin{equation}
y + 2 r^2 \frac{\sin(y)}{\cos(y) - \cosh(x)} = 0
\end{equation}
The solutions, displayed in figure \ref{fig:contours}, consist of the real line, together with a loop encircling the origin.
Starting from the positive saddle $L_0$, the steepest descent contour leaves the real axis along the loop and intersects the negative saddle $-L_0$.
Starting from $-L_0$ the steepest descent contour covers the negative real axis.

To carry out the integral defined by \eqref{pr} we introduce a substitution $w = \coth(L/2)$.
Under this substitution we obtain
\begin{align} \label{wintegral}
\psi(r) &= - 2 \mu_L \int dw (w+1)^{\frac{1}{4G} + \frac{b}{2} - 1}
(w-1)^{-\frac{1}{4G} + \frac{b}{2} - 1} e^{\frac{r^2}{2 G} w}.
\end{align}
The saddle points at $\pm L_0$ are mapped to the points $\pm w_0$ where $w_0 = \sqrt{1+1/r^2}$.
The positive $L$-axis is mapped to $w > 1$, while the negative $L$ axis is mapped to $w < -1$.
These two lines are separated by a branch cut for $-1 < w < 1$; crossing the branch cut shifts the imaginary part of $L$.

We can carry out the integral along either steepest descent contour using known integral representations of the confluent hypergeometric functions (\cite{NIST:DLMF}, \S 13.4)
\begin{align}
I_1 &:= \int_{-1}^{-\infty} dw \; (w+1)^{a-1} (w-1)^{-a+b-1} e^{\frac{z}{2} w} = - \Gamma(a) \, 2^{b-1}  e^{-z/2}  U(a,b,z), \\
I_2 &:= \oint_\gamma dw \; (w+1)^{a-1} (w-1)^{-a+b-1} e^{\frac{z}{2} w} = 2 \pi i \frac{\Gamma(a)}{\Gamma(1+a-b)} 2^{b-1} e^{-z/2} \mathbf{M}(a,b,z). \label{I2}
\end{align}
We have reintroduced the variables $a,z$ from subsection \ref{subsection:exact}.
The curve $\gamma$ encircles the interval $(-1,1)$ clockwise.
We see that these integrals give solutions to the Wheeler-DeWitt equation, as required.
$I_1$ is real, while $I_2$ is imaginary.

The simplest solution to the conformal mode problem is simply to rotate the contour to the negative real axis, resulting in a convergent integral \cite{Gibbons:1978ac}.
In \cite{Bautista:2019jau}, it was shown how conformal bootstrap can be used to resolve the ambiguities associated to the analytic continuation of the integration over the Weyl mode in two space time dimensions. 
However, we see here that the contour does not pass through the saddle point at $L_0$, so gives a partition function which is exponentially suppressed.
The resulting entropy is negative: it is given by $-L_0/4G$ in the classical limit.
Thus the naive resolution to the conformal mode problem gives an unphysical result in our application.

Instead, we can consider a contour passing through the saddle point at $L_0$.
This contour terminates on another saddle point, the one at $-L_0$.
In cases such as this when a steepest descent curve intersects another saddle point, the saddle point is said to be on a Stokes line.
We must decide how to extend the contour past the other saddle point.
The standard method to deal with this case is to analytically continue the parameters of the problem to complex numbers: for example, by giving $G$ a small imaginary part, $G \to G \pm i \epsilon$.
When we do this, the steepest descent contour passing through $L_0$ slightly misses the saddle point at $-L_0$ and continues close to the negative real axis.
As we take $\epsilon \to 0$ the contour becomes a union of the loop encircling the origin and the negative real axis.

However, depending on the sign of the imaginary part of $G$, the contour will traverse $(-\infty,0)$ in either the positive or negative direction.
Thus the real part of $\psi(r)$ will be discontinuous as a function of the complexified $G$.
A natural prescription in this case is to take the average of the two results \cite{Aniceto:2013fka}.
This cancels out the contribution from the negative real axis, and the result is proportional to the loop integral \eqref{I2}.
While this gives a purely imaginary integral, the result for $\psi(r)$ can be made real by choosing an imaginary value for the measure factor $\mu_L$.

We can further choose the measure factor $\mu_L$ so that the partition function satisfies $Z(r) \to 0$ as $r \to 0$.
The result is
\begin{equation}
    \psi(r) = e^{-r^2/2G} M\left(\frac{1}{4G} + \frac{b}{2},b, \frac{r^2}{2G} \right).
\end{equation}
This is the same as the exact solution obtained in section \ref{subsection:exact}.

We note that other choices of contour are possible. which could also include contributions from complex saddle points.
Each of these contributions comes with a nontrivial phase, but they can be summed over to give a real result.
We note, for example, that by choosing the Pochhammer contour in carrying out the integral \eqref{wintegral}, we obtain a result proportional to $M(a,b,z)$ c.f. (\cite{NIST:DLMF}, 13.4.11).

\section{Entanglement Entropy}
\label{section:entanglement}

We now consider entanglement entropy in the $T \overline{T}$-deformed theory.
Specifically we will consider the setup of Ref.~\cite{Donnelly:2018bef}, where the conformal field theory lives on a sphere and the entangling surface consists of two antipodal points.
This corresponds to the entanglement entropy of the Hartle-Hawking state of the CFT, viewed as a state on a circle divided into two semicircles.
Equivalently, it is the de Sitter entropy of the CFT.

This calculation was carried out in Ref.~\cite{Donnelly:2018bef} in the strict large $c$ limit.
This corresponds to the classical limit of the bulk gravity theory, and in this limit the result reproduces the Ryu-Takayanagi formula \cite{Ryu:2006bv}.

However, our result also includes $1/c$ corrections, which correspond to quantum corrections in the bulk.
These corrections to the entanglement entropy are conjectured to capture entanglement of the bulk fields across the minimal surface \cite{Faulkner:2013ana}:
\begin{equation}
    S = \frac{\langle L \rangle}{4G} + S_\text{bulk} + O(1/c).
\end{equation}
In the present case the bulk theory is pure gravity and has no local degrees of freedom, so it is not clear exactly what bulk degrees of freedom could be responsible for $S_\text{bulk}$.

Higher-order corrections in the $1/c$ expansion generically are expected to deform the location of the extremal surface to a \emph{quantum extremal surface} \cite{Engelhardt:2014gca}.
In our case, the location of the bulk surface is fixed by rotational symmetry, so the quantum extremal surface coincides with the extremal surface at all orders: a geodesic through the center of the bulk.
In this case, the higher order corrections to the entanglement entropy of the boundary are higher order corrections in the semiclassical expansion of $\frac{\langle L \rangle}{4G} + S_\text{bulk}$ about the classical minimal surface.

Even when a theory has no local degrees of freedom, it still has an entanglement entropy.
The best known example is Chern-Simons theory, 
where the entanglement comes from edge modes localized on the entangling surface
\cite{Wen:2016snr,Fliss:2017wop,Wong:2017pdm}.
Since 3D gravity is closely related to Chern-Simons theory \cite{Witten:1988hc}, we might expect that the bulk entanglement entropy has a similar description in terms of edge modes.

To precisely describe entanglement entropy in terms of  bulk gravity, we need a description of the gravitational edge modes and their multiplicities.
It is not known how to do this for general relativity, but some aspects of the problem at the classical level were worked out in Ref.~\cite{Donnelly:2016auv}.
In 3D gravity there are a number of more specific proposals
, see e.g. \cite{McGough:2013gka,Geiller:2017xad,Wieland:2017cmf}.
We expect the area to play a preferred role in the edge modes for gravity, based on the algebra and generators and also analogy with the Ryu-Takayanagi formula \cite{Harlow:2016vwg, Lin:2017uzr}.
In 3D gravity, the total length is the only invariant of the intrinsic geometry of the entangling surface.
Moreover, calculations in holography \cite{Dong:2018seb, Akers:2018fow} show that the entanglement spectrum is flat for fixed area states, at least to leading order in the $1/N$ expansion.
This suggests that the gravitational edge modes in 3D are labelled by the length of the entangling surface, with a multiplicity given by $\exp\left( \frac{L}{4G} \right)$ in the classical limit.

We will find some evidence for this picture, namely that the corrections to the entropy of the boundary theory capture fluctuations of the length of the bulk geodesic.
However, this interpretation relies on choosing a real contour for the gravitational path integral.
The interpretation of the entropy as fluctuations of the bulk length operator is obscured when the length is continued to complex values.
We will comment on this further in the discussion.

\subsection{Entropy for antipodal points on the sphere}

We first briefly review the calculation of the entanglement entropy in the special case of antipodal points, and its relation to the sphere partition function.
We are essentially repeating the argument of Ref.~\cite{Donnelly:2018bef}.

The entropy is computed through the replica trick, which first involves evaluating the partition function on an $n$-sheeted branched cover of the sphere, where the branch points are at the entangling surface. 
The line element on such a space is:
\begin{equation}
    \dd s^{2}=r^{2}[\dd \theta^{2}+n^{2}\sin^{2}(\theta) \dd \phi^{2}].
\end{equation}

The entanglement entropy is then given by the formula:
\begin{equation}
S=\left(1- n \dv{n} \right)\log Z \eval_{n=1}.
\end{equation}
In the absence of rotational symmetry, the partition function must be analytically continued to a neighbourhood of $n=1$.
In the present situation rotational symmetry allows us to vary $n$ continuously.

Under infinitesimal variations of $n$, the generating functional responds as follows:
\begin{equation}
-\int \textrm{d}^{2}x \sqrt{\gamma}\ev{T^{\phi}_{~\phi}}=\dv{\log Z}{n}\eval_{n=1}.
\end{equation}
In the limit $n\rightarrow 1$, the full spherical symmetry is re-instated, so the one point function of the energy momentum tensor is isotropic $\ev{T^{ij}}=\alpha g^{ij}$. This in particular tells us that $\ev{T^{\phi}_{~\phi}}=\frac{1}{2}\ev{T^{i}_{~i}}$. This means that the von Neumann entropy in the situation at hand can be computed through the formula: 
\begin{equation} \label{sphere-entropy}
 S = \left( 1 - \frac{r}{2} \dv{r} \right) \log Z.
\end{equation}
Here $Z$ is the sphere partition function without a conical singularity.

Thus we can obtain the entanglement entropy directly from the sphere partition function.
The formula \eqref{sphere-entropy} is the usual thermodynamic formula for the entropy in terms of the partition function, but with $r^2$ playing the role of inverse temperature $\beta$.
We will return to this point shortly.

Note that the counterterm introduces a shift in the partition function of the form $\log Z \to \log Z + \alpha r^2$ where $\alpha$ is constant.
This shift of the partition function does not change the entanglement entropy \eqref{sphere-entropy}, so the counterterm drops out of the calculation of the entropy.

\subsection{Loop expansion of the entropy}

It is straightforward to apply this formula to the partition function in the WKB approximation, yielding
\begin{equation}
S = \frac{1}{2G} \sinh^{-1} \left( \frac{r}{\ell} \right) - \frac{1}{2} (2b - 1) \log(r) - \frac{1}{4} \log(1 + r^2) + \frac{2b-1}{4} + \frac{1}{4} \frac{r}{1 + r^2}.
\end{equation}
The first term, propotional to $1/G$, is the classical term $\frac{L_0}{4G}$.
The remaining terms are a one-loop quantum correction.
We recall that this solution is ambiguous up to the addition of a constant, which can be fixed by the choice of boundary conditions.

Another limit of interest is one where $r\gg \ell$.
In the large $r$ limit, we find
\begin{equation}
S = \left(\frac{\ell}{2G} - b \right) \log r + O(r^0) =  \left( \frac{c}{3} - b \right) \log(r) + O(r^0).
\end{equation}
This gives a correction to the CFT result, which is small if we hold $b$ fixed in the large $c$ limit.

\subsection{Finite radius FLM corrections}

We recall that the partition function takes the form
\begin{equation} \label{thermal}
    Z(r) = \int \, \mu_L dL \, \sinh \left( \frac{L}{2} \right)^{-b}  \exp \left[\frac{L}{4G} + \frac{r^2}{2G} \coth \left( \frac{L}{2} \right) \right],
\end{equation}
up to a counterterm which does not affect the entropy.
Provisionally, we will treat \eqref{thermal} as though it were a convergent real integral, returning to the issues of the conformal mode in due course.

We note that \eqref{thermal} resembles a canonical thermal partition function in which the states are labelled by $L$, with density of states $dn$ and energy $E$,
\begin{equation}
    Z = \int dn(L) e^{-\beta E(L) }.
\end{equation}
In this equation we identify $r^2$ with the inverse temperature $\beta$; this is consistent with the formula \eqref{sphere-entropy} for the entropy.
The density of states and energy can be read off from \eqref{thermal} as:
\begin{align}
    dn(L) &= \mu_L dL \, \sinh \left( \frac{L}{2} \right)^{-b} \exp \left[ \frac{L}{4G} \right], \\
    E(L) &= - \frac{1}{2G} \coth \left( \frac{L}{2} \right).
\end{align}
Including the counterterm simply shifts $E(L)$ by a constant.

We can now relate the entropy calculated by the sphere trick to fluctuations of the length $L$.
The distribution over lengths implied by this canonical distribution is given by the measure
\begin{equation}
    d\rho(L) = \frac{1}{Z} e^{- \beta E(L)} dn(L).
\end{equation}
Since $L$ is a continuous parameter, the entropy of the distribution $d\rho$ is not invariant under reparametrizations of $L$.
Instead, one should consider the relative entropy $S(\rho \Vert \sigma)$ where $\sigma$ is a reference distribution:
\begin{equation}
    S(\rho \Vert \sigma) = \int d\rho(L) \log \left( \frac{d \rho(L)}{d \sigma(L)} \right).
\end{equation}
This quantity is invariant under reparametrizations when both $d \rho$ and $d \sigma$ transform as measures.
Note that the sign is opposite from the one appearing in the entropy, $S = - \sum p \log p$.

Having put the partition function into the canonical form, we can straightforwardly calculate the entropy.
It takes the suggestive form
\begin{equation} \label{entropy}
    S = \frac{\langle L \rangle}{4G} - S(\rho \Vert \sigma),
\end{equation}
where $\langle L \rangle$ denotes the expectation value in the distribution $\rho$.
The reference distribution $\sigma$ is defined by
\begin{equation}
    d \sigma(L) = \mu_L dL \sinh \left( \frac{L}{2} \right)^{-b}.
\end{equation}
This suggests an interpretation in which the gravitational edge modes are labelled by the length $L$ of the bulk geodesic.
The number of distinct eigenvalues of $L$ is given by the measure $d \sigma(L)$, while the degeneracy of the eigenvalues is given by $e^{L/4G}$.
This would appear to give a realization of the Faulkner-Lewkowycz-Maldacena proposal \cite{Faulkner:2013ana} in which the bulk entanglement entropy can be understood as arising from gravitational edge modes labelled by the length. 

Unfortunately, such a nice interpretation seems to be precluded by the conformal mode problem.
When the contour for the $L$ integral is complex, the interpretation of the states being labelled by a real geometric length is not available.
We do not know whether there is any interpretation of the entropy analogous to \eqref{entropy} when $L$ is allowed to be complex.

\section{Discussion}

We have studied the $T \overline{T}$-deformation of 2D conformal field theories, showing that the equation for the trace anomaly maps directly to the Wheeler-DeWitt equation of 3D gravity with a negative cosmological constant.
The diffeomorphism invariance that emerges from this construction is a powerful tool, and we have shown that it can be used to determine the partition function on the sphere, and hence the entanglement entropy for an entangling surface consisting of antipodal points on the sphere.
The resulting entropy captures the length of a bulk geodesic, together with both perturbative and nonperturbative quantum gravity corrections --- the latter are conjectured to be dual to bulk entanglement entropy, and our calculation suggests an interpretation as entropy of gravitational edge modes labelled by the fluctuating length of the bulk geodesic.

While the $T \overline{T}$ deformation inherits many advantages from metric quantum gravity, it also inherits some disadvantages.
In particular, we have found that the sphere partition function suffers from a version of the conformal mode problem.
While we can resolve this problem by a judicious choice of contour, this procedure requires integrating over complex geometries which obscures the interpretation of the entropy as a measure of geometric fluctuations of the bulk gravity theory.
We leave it as an open puzzle whether the entropy we calculate can be given a state counting interpretation in terms of bulk quantities.

The entanglement entropy that we have computed was obtained through the replica trick specialized to the case where the entangling surface consists of antipodal points on the sphere. 
When applied to a local quantum field theory on $S^{2}$, this entropy has a statistical interpretation, in terms of micro-state counting of CFT states in the two-dimensional de Sitter static patch. 
In the case at hand however, the theory we are considering is known not to be strictly local, and can be shown to arise from integrating out fluctuations of two dimensional geometry \cite{Dubovsky:2017cnj,Dubovsky:2018bmo,Cardy:2019qao}.
This means that it isn't entirely clear whether the replica trick does indeed compute an entropy in the traditional sense.
One possible test is to calculate the full entanglement spectrum, which can be obtained from the R\'enyi entropies.

Another interesting result to mention in this vein is that of \cite{Freidel:2008sh}, where an integral transform between a CFT partition function and a solution to the bulk radial Wheeler-DeWitt equation was presented. The integration was over the frame fields of the geometry on which the CFT lived and it was weighted by a Gaussian kernel. This simple formula can be taken to be a direct definition of the $T\bar{T}$ deformed partition function at finite values of the deformation parameter. This perspective will be elucidated in upcoming work \cite{Mazenc:2019cfg}. 

Ref.~\cite{Donnelly:2018bef} found some non-analyticity in the spectrum of R\'enyi entropies in the $c \to \infty$ limit.
It would be interesting to see whether this non-analyticity is also present in the theory at finite $c$, or if it is a consequence of the infinite $c$ limit.

Calculating R\'enyi entropy requires evaluating the partition function on a manifold with conical defect, so the gauge-fixing procedure used in section \ref{section:s2} would have to be generalized to a larger reduced phase space than the one considered here.
The ability to evaluate the partition function for more general geometries would also enable the a calculation of entanglement entropy beyond the case of antipodal points; see \cite{Grieninger:2019zts} for some work at the classical level.

For general boundary geometries, it is known that fixed metric boundary conditions for Euclidean general relativity are not elliptic \cite{Witten:2018lgb}. 
We do not yet know the implications of this result for calculating entanglement entropy beyond the spherically symmetry sector.
In particular, does this become an obstacle when computing the bulk entropy in perturbative Euclidean quantum gravity? 
We note that our results contained an independent constant $b$, which arose as an ordering ambiguity in the quantization of the Hamiltonian constraint.
It would be interesting to know whether the value of this constant could be fixed on physical grounds.
We note that if we divide the Wheeler-DeWitt operator into kinetic and potential terms $K = \tfrac{1}{2}(\partial_r^2 + (2b-1) r^{-1} \partial_r)$, $V = \tfrac{1}{2} r^2$ and introduce the dilatation operator $D = r \partial_r$ we find a modified $\mathfrak{sl}(2,\mathbb{R})$ algebra:
$$[D,K] = -2K, \qquad [D,V] = 2V, \qquad [K,V] = D + b.$$
A similar modification appears in the quantization of Jackiw-Teitelboim gravity \cite{Iliesiu:2019xuh}, which related to 3D gravity by dimensional reduction.
This suggests that the the constant $b$ may be related to the breaking of $\mathrm{SL}(2,\mathbb{R})$ symmetry by quantum effects.

Our results bear some similarity to those of \cite{Caputa:2018asc}, where the $1/N$ resummed sphere partition function of ABJM theory in three space-time dimensions was computed via a bulk minisuperspace path integral calculation. 
However, the justification of the minisuperspace approximation in their work was different from ours and was due to supersymmetric localization.
Nevertheless, an interesting question for future investigation is whether the $1/N$-corrected entanglement entropy of the Hartle--Hawking state can be obtained in ABJM theory through the methods used in this article. 

Another interesting generalization of this work would be to apply it in the context of the dS/dS correspondence put forward in \cite{Gorbenko:2018oov}. 
It was noted there that the boundary theory that inhabits a finite, time like dS$_{2}$ boundary of dS$_{3}$ is given by the deformation of a CFT by first the $T\overline{T}$ operator and then turning on the boundary cosmological constant operator. 
This defines a flow equation which, under the change of variables used to identify bulk and boundary quantities in this article, becomes the radial Hamiltonian constraint equation with a \emph{positive} cosmological constant. 
The phase space reduction and subsequent quantization of that theory should proceed in a manner very similar to what has been presented in this article. 
Doing so would allow one to ask interesting questions about finite volume regions in de Sitter quantum gravity. 

\section*{Acknowledgments}
We are grateful to the organizers of the 2019 Perimeter Scholars International winter school where this work was initiated.
WD thanks Hal Haggard, Etera Livine, and Ricardo Schiappa for useful conversations. VS would like to thank Steven Carlip,  Vincent Moncrief and Zohar Komargodski for helpful email exchanges and Alexander Zamolodchikov, Ronak Soni and Edward Mazenc for illuminating discussions. 
This research was supported in part by Perimeter Institute for Theoretical Physics.
Research at Perimeter Institute is supported by the Government of Canada through the Department of Innovation, Science and Economic Development Canada and by the Province of Ontario through the Ministry of Research, Innovation and Science. 
AP would also like to thank the São Paulo Research Foundation (FAPESP) for partial financial support through grant 2017/11341-4.

\bibliographystyle{utphys}
\bibliography{ttbar}

\end{document}